\documentclass{iaus}
\usepackage{graphicx}

\title[Magnetic fields, winds and X-rays of massive stars in the Orion Nebula Cluster ] 
{Magnetic fields, winds and X-rays of massive stars in the Orion Nebula Cluster }

\author[Petit et al.]   
{V. Petit$^1$, G.~A. Wade$^2$, E. Alecian$^3$, L. Drissen$^4$, T. Montmerle$^3$ \and A. ud-Doula$^5$}
\affiliation{$^1$Dept. of Geology \& Astronomy, West Chester University, West Chester, PA 19383, USA  \\ email: {\tt VPetit@wcupa.edu} \\
[\affilskip]
$^2$Dept. of Physics, Royal Military College of Canada, Kingston, K7K 7B4, Canada\\
[\affilskip]
$^3$Laboratoire d'Astrophysique de Grenoble, F-38041 Grenoble Cédex 9, France \\
[\affilskip]
$^4$D\'ept. de Physique, Universit\'e Laval, Qu\'ebec, Canada \\
[\affilskip]
$^5$ Penn State Worthington Scranton, Dunmore PA 18512, USA  \\
}

\pubyear{2010}
\volume{272}  
\pagerange{119--126}
\jname{Active OB stars: structure, evolution, mass loss, and critical limits}
\editors{C. Neiner, G. Wade, G. Meynet \& G. Peters, eds.}
\begin{document}

\maketitle

\begin{abstract}

In some massive stars, magnetic fields are thought to confine the outflowing radiatively-driven wind. Although theoretical models and MHD simulations are able to illustrate the dynamics of such a magnetized wind, the impact of this wind-field interaction on the observable properties of a magnetic star - X-ray emission, photometric and spectral variability - is still unclear. The aim of this study is to examine the relationship between magnetism, stellar winds and X-ray emission of OB stars, by providing empirical observations and confronting theory. In conjunction with the COUP survey of the Orion Nebula Cluster, we carried out spectropolarimatric ESPaDOnS observations to determine the magnetic properties of massive OB stars of this cluster. 
\keywords{stars: early-type, stars: magnetic fields, X-rays: stars, techniques: polarimetric}
\end{abstract}

\firstsection 

\section{Introduction}

The Chandra Orion Ultradeep Project (COUP) was dedicated to observe the Orion Nebula Cluster (ONC) in X-rays. The OBA sample (20 stars) was studied with the goal of disentangling the respective roles of winds and magnetic fields in producing X-rays (Stelzer et al. 2005). 
The production of X-rays by radiative shocks (Lucy \& White 1980, Owocki \& Cohen 1999) should be the dominant mechanism for the subsample of 9 O to early-B stars which have «strong winds». However, aside from 2 of those stars, all targets showed X-ray intensity and/or variability which were inconsistent with the small shock model predictions. 
We have undertaken a study with ESPaDOnS to explore the role of magnetic fields in producing this diversity of X-ray behaviours. 

Eight stars of the COUP «strong winds» OB subsample were observed with the echelle spectropolarimeter ESPaDOnS at CFHT. 
The mean Stokes I and V profiles were extracted with the Least Squares Deconvolution technique (LSD) of Donati et al. (1997), which allows the use of many lines to increase the level of detection of a magnetic Stokes V signature.  Formal signal detection was achieved for 3 stars: $\theta^1$\,Ori\,C (for which a field has already been detected by Donati et al. 2002), LP\,Ori (HD\,36982) and NU\,Ori (HD\,37061). 

With foreknowledge of the shape of an expected deviation, we can add some information to the formal $\chi^2$ statistics. It has been shown that with multiple noisy observations, it is possible to pick up an underlying signal by computing the odds ratios of the no magnetic field model ($M_0$) to a magnetic model ($M_1$), in a Bayesian framework (Petit et al. in prep).
As the exact rotation phases of our observations are not known, we used the method described by Petit et al. (2008), which compares the observed Stokes V profiles to a rotation independent, dipolar oblique rotator model.  
As can be seen from the computed odds (Table \ref{tab_res}) in the case of the detected stars, the magnetic oblique rotator model is favoured by many orders of magnitude ( $\log(M_0/M_1) < 0$ ). For the non-detected stars, any improvement of the fit to the data achieved by assuming a magnetic field is not sufficient to justify employing this more complex magnetic model.

By performing a Bayesian parameter estimation for the dipole model, we can obtain the probability density function marginalised for the dipole strength, and put constraints on the values admissible by our observations ($B_\mathrm{pole}$ in Table \ref{tab_res}).

\begin{table}
  \begin{center}
  \caption{Odds ratios, surface dipolar strength and magnetic wind confinement of ONC stars.}
  \label{tab_res}

\renewcommand{\arraystretch}{1.25}
 {\scriptsize
\begin{tabular}{l c c c c c}
\hline
\multicolumn{1}{c}{ID}		& HD 	&Spec Type			&$\log(M_0/M_1)$	& $B_\mathrm{pole}~^1 [G]$ 					& $B_\mathrm{pole}(\eta_\star=1.0)$ [G] \\
\hline
$\theta^1$\,Ori\,C 	& 37022 	& O7\,V\,$^\dagger$		& \textbf{-110}		&{$1785\stackrel{+494}{_{-652}}$}	& 268 \\ 
$\theta^2$\,Ori\,A 	& 37041	& O9.5\,V\,$^\dagger$	& 0.38			&$<118$ 				& 118 \\ 
$\theta^1$\,Ori\,A	& 37020	& B0.5\,V\,$^\dagger$	& 0.51			&$<57$ 				& 63 \\ 
$\theta^1$\,Ori\,B 	& 37023	& B0.5\,V\,$^\dagger$	& 0.38			&$<79$ 				& 77 \\ 
NU\,Ori			& 37061	& B0.5\,V\,$^\dagger$	& \textbf{-15}		&$465\stackrel{+116}{_{-179}}$		& 66 \\ 
$\theta^2$\,Ori\,B 	& 37042	& B0.5\,V				& 0.25			&$<103$ 				& 45 \\ 
LP\,Ori 			& 36982	&B1-2\,V				& \textbf{-37}		& $1\,020\stackrel{+199}{_{-302}}$	& 15 \\ 
JW\,660 			&		&B3\,V\,$^\dagger$		& 0.14			&$<1\,287$ 				& 16 \\ 
\hline
\end{tabular}
}
 \end{center}
\vspace{1mm}
 \scriptsize{
 $^1$ Median of the posterior probability density marginalised for $B_\mathrm{pole}$ and 68.3\% credible region for magnetic stars for detected stars, upper limit of the 95.4\% credible region for the non-detections.
\\$^\dagger$ Confirmed or suspected binaries.}
\end{table}

\section{Wind confinement}

According to our observations, the 3 magnetic massive stars of the ONC have fields that should be strong enough to dynamically influence their stellar winds at a significant level (see the minimum field required for confinement in Table \ref{tab_res}). However, this field-wind interaction is not reflected in any systematic way in the X-ray properties of these stars. Furthermore, no fields strong enough to dynamically influence the wind are found in other ONC massive stars that Stelzer et al. (2005) considered to be ``prime candidates'' for magnetism.
From this we conclude that X-ray variability, intensity and hardness enhancement are not systematically correlated with the presence of a magnetic field. 

More detailed studies of the field geometries of these magnetic stars will serve as inputs to new models (Townsend et al. 2007) and 3D MHD simulations of magnetic wind confinement (e.g. ud-Doula et al. 2008), to better understand the mechanisms that lead to this variety of X-ray properties.

\end{document}